\documentclass[11pt]{article}
\pdfoutput=1
\usepackage[nosort]{cite}
\usepackage[dvipsnames]{xcolor}
 
\usepackage{epsfig}
\usepackage{booktabs}
\usepackage{amsfonts}
\usepackage{amscd}
\usepackage{latexsym}
\usepackage{dsfont}
\usepackage{amsmath,amssymb}
\usepackage{bm}
\usepackage{verbatim}
\usepackage{setspace}
\usepackage{color}
\usepackage{fancyhdr}
\usepackage{cite}
\usepackage{hyperref}
\usepackage{tikz}
\usepackage{slashed}
\usepackage{multirow}
\usepackage{amsthm}
\newtheorem*{theorem*}{Theorem}

\usetikzlibrary{topaths}
\usetikzlibrary{decorations}
\usetikzlibrary{decorations.pathmorphing}
\usetikzlibrary{arrows, decorations.markings, cd}
\usetikzlibrary{calc, cd, decorations.pathmorphing}
\tikzset{
->-/.style args={#1rotate#2}{decoration={markings, mark=at position #1 with {\arrow[scale=1.5,rotate = #2 ]{stealth}}}, postaction={decorate}}
}
\usetikzlibrary{shapes.geometric}
\usetikzlibrary{knots}
\usepackage{tikz-3dplot}
\usetikzlibrary{positioning}

\usepackage{draft}
\usepackage{hyperref}
\usepackage{graphicx,subfig}
\usepackage{cite}
\usepackage{mciteplus}
\usepackage{bbm}
\usepackage{braket}
\usepackage[most]{tcolorbox}

\numberwithin{equation}{section}

\def\({\left(}
\def\){\right)}

\newcommand{\g}{\gamma}
\newcommand{\TD}{\Theta_{\text{D}}}
\newcommand{\hg}{\hat\gamma_5}
\newcommand{\hP}{\hat P}

\newcommand{\im}{\operatorname{im}}

\begin{document}

\hfill MIT-CTP/6151

\title{On reflection positivity of lattice chiral fermions}

 \author{Zhiyao Lu and Shu-Heng Shao\\[2mm]
 {\small \textit{Center for Theoretical Physics --- A Leinweber Institute, Massachusetts Institute of Technology,\\
 Cambridge, MA 02139, USA}}}

\begin{abstract}
\noindent

We point out that the overlap Weyl fermion, the spacetime lattice
realization of a Weyl fermion underlying L\"uscher's 
construction of lattice chiral gauge theories, is not reflection
positive in two and four dimensions at finite lattice spacing.
Consequently, the standard Osterwalder--Schrader reconstruction
does not yield a Hilbert space with a self-adjoint Hamiltonian.
In two dimensions, we further show that the overlap Weyl fermion
cannot be cut open: the lattice action becomes nonlocal
in the presence of a boundary, as expected from the gravitational
anomaly of the continuum field theory.
By contrast, the overlap Dirac fermion is known to be reflection positive.
However, its compact axial symmetry is not invariant under the
reflection, whereas a reflection-invariant axial symmetry is
noncompact, and both are part of an Onsager algebra.

\end{abstract}

\newpage
\tableofcontents

\section{Introduction}

Constructing lattice models with exact chiral global symmetries, and ultimately lattice chiral gauge theories, is a long-standing problem of both practical and conceptual importance. 
A celebrated construction starts from the observation of Ginsparg and Wilson that a lattice Dirac operator can retain a remnant of chiral symmetry if it obeys a certain relation \cite{Ginsparg:1981bj}. Neuberger later found a closed-form solution, the overlap Dirac operator \cite{Neuberger:1997fp,Neuberger:1998wv}, from the formalism he developed with Narayanan \cite{Narayanan:1993zzh,Narayanan:1993sk,Narayanan:1994gw}, which was in turn motivated by domain-wall fermions \cite{Kaplan:1992bt,Shamir:1993zy,Furman:1994ky}. L\"uscher then showed that the Ginsparg--Wilson relation implies a modified chiral symmetry of the lattice action, which is exact, compact,  and locality-preserving (with exponential tails) \cite{Luscher:1998pqa}. 
The resulting model, known as the \textit{overlap Dirac fermion}, describes a single Dirac fermion in the continuum limit with no doublers.

The same construction also yields a lattice Weyl fermion, obtained by restricting the fermion fields to a fixed chirality with respect to the modified chiral symmetry \cite{Narayanan:1998uu,Luscher:1998du}. 
The resulting model, known as the \textit{overlap Weyl fermion}, is the basis of L\"uscher's construction of abelian chiral gauge theories \cite{Luscher:1998du}. It was extended to nonabelian theories in perturbation theory \cite{Luscher:1999un,Luscher:2000zd}, and recently nonperturbatively by Craig \cite{Craig:2026soq}; see also \cite{Suzuki:2000ku,Kadoh:2007wz,Kadoh:2007xb,Kikukawa:2017ngf} and the reviews \cite{Niedermayer:1998bi,Luscher:2000hn,Kaplan:2009yg}. A different approach keeps the fermions in the continuum and discretizes only the gauge fields \cite{Hsu:2026gmi,Witten:2026ezr} (see \cite{Misumi:2026ham} for numerical tests). 

The overlap Weyl fermion without gauge fields is a spacetime lattice model with an exact $\mathrm{U(1)}$ global symmetry whose continuum limit is a single Weyl fermion. 
No such model exists in the Hamiltonian lattice formalism in $1+1$ and $3+1$ dimensions.\footnote{
If the U(1) global symmetry is not imposed, a single Weyl fermion
in 3+1 dimensions can be realized in a Hamiltonian lattice model
\cite{Nielsen:1981xu,Gioia:2025bhl}.
By contrast, a single Weyl or Majorana-Weyl fermion in 1+1 dimensions
is impossible to realize in a local Hamiltonian
lattice model, even without imposing any global symmetry,
due to its purely gravitational anomaly \cite{Jones:2026gmd}.
} 
For free fermions this is the Nielsen--Ninomiya theorem \cite{Nielsen:1980rz,Nielsen:1981xu,Nielsen:1981hk,Friedan:1982nk}. 
For interacting fermions it was shown by Shamir under certain assumptions on the low-energy spectrum \cite{Shamir:1993bi}, and more recently for anomalous $\mathrm{U(1)}$ symmetries implemented by locality-preserving unitaries on finite-dimensional local Hilbert spaces \cite{Fidkowski:2023sif,Kapustin:2024rrm,Liu:2026atf}.\footnote{Note that these no-go theorems do not apply to Hamiltonian lattice models with infinite-dimensional local Hilbert spaces (e.g., rotors or the modified Villain models), where anomalous continuous global symmetries can be realized exactly \cite{Cheng:2022sgb,Fazza:2022fss,Fidkowski:2025rsq,Thorngren:2026ydw,Seifnashri:2026ema,Lu:2026jnq,Dharanikota:2026vik,Lu:2026itw,Jones:2026gmd,Shao:2026ncb,Lootens:2026kcg} (see also \cite{Gioia:2026wdl}).} 
The analogous no-go theorems for Euclidean spacetime lattice models \cite{Karsten:1981gd,Karsten:1980wd,Baig:2026zgb} do not apply to the overlap Weyl fermion either, because its fermion fields are not independent Grassmann variables.

How can a spacetime lattice model achieve what a Hamiltonian lattice model cannot? 
A Euclidean spacetime lattice model leads to a Hilbert space and a positive transfer matrix, and hence to a Hamiltonian lattice model, if it is reflection positive \cite{Osterwalder:1973dx,Osterwalder:1974tc,Osterwalder:1977pc,Luscher:1976ms}. 
On the infinite lattice, in both two and four spacetime dimensions, we show that the free overlap Weyl fermion is not reflection positive with respect to any reflection that respects locality. Reflection positivity is known only for a restricted set of observables \cite{Kikukawa:2010gq}. 
The obstruction is topological. 
The fermion fields of the overlap Weyl fermion form a vector bundle over the Brillouin torus with a nonzero Chern number. 
If the model were reflection positive with respect to some reflection satisfying a set of  assumptions from locality and symmetries, this bundle would have a nowhere-vanishing global section, which is impossible.
This bundle is the pullback of the $\mathrm{U(1)}$ monopole bundle on $S^2$ in two dimensions, and of the $\mathrm{SU(2)}$ instanton bundle on $S^4$ in four dimensions.

Having said that, reflection positivity at finite lattice spacing is not necessary for a lattice model to be useful. 
Many lattice QCD actions used in practice, such as improved gauge actions, are not reflection positive \cite{Luscher:1984is}, yet they produce numerical results that are successfully compared with experiment. 
Our result also does not rule out that reflection positivity emerges in the continuum limit of the overlap Weyl fermion and of L\"uscher's chiral gauge theories. 
However, Monte Carlo sampling of the Euclidean path integrals of these theories suffers from a sign problem, and it is not clear whether they can be simulated numerically. 
Their value at this point lies mostly in the conceptual question of whether chiral gauge theories admit a nonperturbative lattice definition. 
We therefore believe that their structural properties, such as reflection positivity, warrant a discussion. 
In particular, our result obstructs the standard Osterwalder--Schrader
reconstruction of a Hilbert space for the free overlap Weyl fermion
at finite lattice spacing. 
This suggests that the chiral gauge theories \cite{Luscher:1998du,Luscher:1999un,Luscher:2000zd,Suzuki:2000ku,Kadoh:2007wz,Kadoh:2007xb,Kikukawa:2017ngf,Craig:2026soq} built upon the overlap Weyl fermion are not reflection positive either at finite lattice spacing. 
Our result says nothing about the constructions of
\cite{Hsu:2026gmi,Witten:2026ezr}, in which the fermions
remain continuum fields.
See  \cite{Creutz:2004ir,Bietenholz:2005rd} for discussions of reflection positivity of lattice gauge fields, an issue emphasized in \cite{Craig:2026soq}.

The contrast between Euclidean spacetime and Hamiltonian lattice models is even sharper in two dimensions, where a Weyl fermion has a gravitational anomaly captured by the chiral central charge $c_L-c_R=1$.
A conventional lattice model can be cut open, which produces a boundary condition for its continuum limit.
A quantum field theory with a pure gravitational anomaly, however, admits no boundary condition. For instance,  a single Weyl fermion carries a net energy flow in one direction, which cannot be absorbed or reflected at a boundary \cite{Cardy:1989ir,Jensen:2017eof,Thorngren:2020yht,Hellerman:2021fla}.
This is why a gravitational anomaly has long been believed to obstruct any local lattice realization. This has been shown recently for Hamiltonian lattice systems, even with infinite-dimensional local Hilbert spaces \cite{Jones:2026gmd}.\footnote{By contrast, a single Weyl fermion can be realized in a 1+1-dimensional Hamiltonian lattice model if one gives up locality of the Hamiltonian \cite{PhysRevD.26.468,Zakharov:2026nly}. See Appendix A of \cite{Shao:2026ncb} for a related model with infinite entanglement entropy, and also \cite{Kaplan:2023pvd} for a Weyl fermion on the boundary of a finite 2+1-dimensional Hamiltonian lattice. }
Nonetheless, the two-dimensional overlap Weyl fermion has a local action on a closed lattice, and one may ask what happens when the lattice is cut open.
We show that it cannot be cut open locally: in the presence of a boundary, the overlap operator decays only as the inverse distance along the boundary, so the lattice action becomes nonlocal.

The overlap Dirac fermion, by contrast, is reflection positive 
\cite{Kikukawa:2010gq}.
However, L\"uscher's compact axial symmetry neither commutes with
the reflection nor preserves the positive-time algebra.
Consequently, the standard Osterwalder--Schrader construction
does not directly yield a unitary realization of this symmetry
on the reconstructed Hilbert space. 
There is another reflection-symmetric axial symmetry, but it is noncompact, generating $\mathbb R$ rather than $\mathrm{U(1)}$. 
This is consistent with the Hamiltonian no-go result for chiral symmetries of lattice fermions in $3+1$ dimensions \cite{Fidkowski:2023sif}. 
It is also reminiscent of the Hamiltonian lattice models in $1+1$ dimensions where the vector and axial charges are both quantized but do not commute, and generate an Onsager algebra \cite{Chatterjee:2024gje}. 
We show that the two axial symmetries of the overlap Dirac fermion, L\"uscher's and its image under the reflection, generate an Onsager algebra as well.

This paper is organized as follows. 
Section~\ref{sec:model} reviews the overlap Dirac and Weyl fermions and introduces the Weyl bundle. 
Section~\ref{sec:rp} discusses reflection positivity, first for the axial symmetry of the overlap Dirac fermion and then for the overlap Weyl fermion. 
Section~\ref{sec:bdy} studies the two-dimensional free overlap Weyl fermion on a lattice with a boundary. 
Appendix~\ref{app:unitary} reviews how a symmetry of a reflection-positive Euclidean lattice model leads to a unitary operator on the reconstructed Hilbert space. 
Appendix~\ref{app:onsager} works out the Onsager algebra of the overlap Dirac fermion. 
Appendix~\ref{sec:Z} reviews the torus partition function of the two-dimensional overlap Weyl fermion and compares it with the continuum.

\section{Weyl fermion on a spacetime lattice}
\label{sec:model}

We work on a $d$-dimensional hypercubic lattice with $d=2$ or $4$.
We set the lattice spacing to 1.
We start with a torus with $L_\mu$ sites in the $\mu$ direction. The total number of sites is $V=\prod_\mu L_\mu$.
(Section~\ref{sec:bundle} and Section~\ref{sec:rp} concern the infinite lattice $\mathbb Z^d$, whose momentum space is the continuous Brillouin torus $T^d$.) 
We have a pair of $2^{d/2}$-component Grassmann variables per lattice site $x$, denoted as $\psi(x)$ and $\bar\psi(x)$. We emphasize that they are independent Grassmann variables and there is no conjugation relation between them.

On the finite torus, we assume boundary conditions $\psi(x+L_\mu\hat\mu)=e^{2\pi i\theta_\mu}\psi(x)$, so the momenta are $p_\mu=\frac{2\pi}{L_\mu}(n_\mu+\theta_\mu)$. $\theta_\mu=0$ ($\theta_\mu=\frac12$) corresponds to periodic (antiperiodic) boundary conditions.
The $2^{d/2}\times 2^{d/2}$ gamma matrices are hermitian and obey $\{\g_\mu,\g_\nu\}=2\delta_{\mu\nu}$. The matrix $\g_5$ is the product of all the gamma matrices, times a phase chosen so that $\g_5$ is hermitian and $\g_5^2=1$. In $d=2$ we choose $\g_1=\sigma_1$, $\g_2=\sigma_2$, so that $\g_5=-i\g_1\g_2=\sigma_3$. In $d=4$ we use
\begin{equation}
\g_k=\begin{pmatrix}0&-i\sigma_k\\ i\sigma_k&0\end{pmatrix}\ (k=1,2,3),\qquad \g_4=\begin{pmatrix}0&1\\1&0\end{pmatrix},\qquad \g_5=\g_1\g_2\g_3\g_4=\begin{pmatrix}1&0\\0&-1\end{pmatrix},
\end{equation}
with $2\times2$ blocks.
The chiral projectors are $P_\pm=\frac12(1\pm\g_5)$.
As a slight abuse of notation, we will often use the same symbol with different arguments for a field or a kernel in position and momentum spaces, i.e., $\psi(x)=V^{-1/2}\sum_pe^{ip\cdot x}\psi(p)$, $\bar\psi(x)=V^{-1/2}\sum_pe^{-ip\cdot x}\bar\psi(p)$, and $K(x-y)=V^{-1}\sum_pe^{ip\cdot(x-y)}K(p)$, so that $\sum_{x,y}\bar\psi(x)K(x-y)\psi(y)=\sum_p\bar\psi(p)K(p)\psi(p)$.

\subsection{The overlap Dirac fermion}

We review the standard construction of the overlap Dirac fermion, which is a free spacetime lattice model for a Dirac fermion \cite{Narayanan:1993zzh,Narayanan:1993sk,Narayanan:1994gw,Neuberger:1997fp,Neuberger:1998wv,Luscher:1998pqa}; see \cite{Niedermayer:1998bi,Luscher:2000hn,Chandrasekharan:2004dph,Kaplan:2009yg} for reviews.
We first define the Wilson operator \cite{Wilson:1975id}
\begin{equation}
D_{\rm w}=\tfrac12\sum_\mu\Big[\g_\mu(\nabla_\mu+\nabla^*_\mu)-\nabla^*_\mu\nabla_\mu\Big],\qquad
D_{\rm w}(p)=i\g_\mu\sin p_\mu+\sum_\mu(1-\cos p_\mu),
\end{equation}
where $\nabla_\mu\psi(x)=\psi(x+\hat\mu)-\psi(x)$, $\nabla^*_\mu\psi(x)=\psi(x)-\psi(x-\hat\mu)$.
The Wilson term gives the $2^d-1$ doublers, the zeros of $\sin p_\mu$ other than $p=0$, a mass. 
The price one pays is the lack of the axial symmetry, because $D_{\rm w}$ does not anticommute with $\g_5$.

To define the overlap fermion, we introduce
\begin{equation}
H_{\rm w}=\g_5(D_{\rm w}-1).
\end{equation}
It is hermitian because 
 $\g_5D_{\rm w}\g_5=D_{\rm w}^\dagger$.\footnote{For a kernel $K(x,y)$, which for each pair of $x,y$ is a matrix on the spinor indices, we define the adjoint kernel $K^\dagger(x,y)=K(y,x)^\dagger$, where $\dagger$ on the right-hand side denotes the conjugate transpose of this matrix, i.e., $\big(K^\dagger(x,y)\big)_{ab}=K_{ba}(y,x)^*$. Throughout, $\dagger$ does not mean the adjoint of an operator acting on a Hilbert space.} 
In momentum space, $H_{\rm w}(p)^2=\omega(p)^2$ with
\begin{equation}
\omega(p)^2=\sum_\mu\sin^2p_\mu+\Big(\sum_\mu(1-\cos p_\mu)-1\Big)^2=1+2\sum_{\mu<\nu}(1-\cos p_\mu)(1-\cos p_\nu)\ \ge1 ,
\label{eq:omega}
\end{equation}
so $|H_{\rm w}|\ge1$ \cite{Neuberger:1997fp,Neuberger:1998wv,Neuberger:1999pz,Adams:1999if}.
Since $H_{\rm w}$ is invertible, the sign function $\mathrm{sgn}(H_\text{w})\equiv H_{\rm w}(H_{\rm w}^2)^{-1/2}$ is well-defined. The overlap operator \cite{Narayanan:1993zzh,Narayanan:1993sk,Narayanan:1994gw,Neuberger:1997fp,Neuberger:1998wv} is
\begin{equation}
D=1+\g_5\,\mathrm{sgn}(H_\text{w}),
\label{eq:overlap}
\end{equation}
and the overlap Dirac fermion action is $\sum_{x,y}\bar\psi(x)D(x-y)\psi(y)$.
This action is not ultra-local, i.e., $D(x-y)$ does not vanish beyond any finite separation \cite{Horvath:1998cm}, but it is local with exponential tails \cite{Hernandez:1998et}.
The overlap Dirac operator $D$ takes a simple form in momentum space \cite{Kikukawa:1998pd},
\begin{equation}
D(p)=(1-\alpha)+i\g_\mu\beta_\mu,
\label{eq:Dp}
\end{equation}
where
\begin{equation}\label{eq:alphabeta}
\alpha(p)=\frac{\sum_\mu\cos p_\mu-(d-1)}{\omega(p)},\qquad \beta_\mu(p)=\frac{\sin p_\mu}{\omega(p)},\qquad \alpha^2+\beta_\mu\beta_\mu=1.
\end{equation}
$D(p)$ vanishes only at $p=0$, where it reduces to the continuum Dirac operator $D\simeq i\g_\mu p_\mu$.
So the overlap Dirac action describes a single massless Dirac fermion.

The advantage of the overlap fermion over the Wilson fermion is that in addition to the $\mathrm{U(1)}_V$ global symmetry implemented by the transformation
\begin{equation}
R_V(\varphi):\ \ \psi\to e^{i\varphi}\psi,\quad \bar\psi\to\bar\psi\,e^{-i\varphi},
\label{eq:V}
\end{equation}
it also has a $\mathrm{U(1)}_A$ axial global symmetry.
To write the axial symmetry, we define the modified chirality and the associated projectors \cite{Luscher:1998pqa,Luscher:1998du,Narayanan:1998uu}
\begin{equation}
\hg=
-\mathrm{sgn}(H_\text{w}),\qquad \hP_\pm=\tfrac12(1\pm\hg).
\label{eq:hg}
\end{equation}
Equivalently, $\hg=\g_5(1-D)$ by \eqref{eq:overlap}. 
Since $\hg$ is hermitian and $(\hg)^2=1$, $\hP_\pm$ are orthogonal projectors.
The axial symmetry transformation is
\begin{equation}
R_A(\varphi):\ \ \psi\to e^{i\varphi\hg}\psi,\quad \bar\psi\to\bar\psi\,e^{i\varphi\g_5}.
\label{eq:A}
\end{equation}
The action is invariant because of the identity $D\hg=-\g_5D$, which follows from the Ginsparg--Wilson relation \cite{Ginsparg:1981bj,Hasenfratz:1998ri}:
\begin{equation}\label{eq:GW}
\g_5D+D\g_5=D\g_5D.
\end{equation}
The $\mathrm{U(1)}_A$ symmetry transformation is not on-site, but it acts locally with exponential tails \cite{Hernandez:1998et}.
Importantly, since $\psi$ and $\bar \psi$ are independent Grassmann variables, this is a valid transformation.
Both $\mathrm{U(1)}_V$ and $\mathrm{U(1)}_A$ are compact (i.e., $R_{V,A}(\varphi+2\pi)=R_{V,A}(\varphi)$) and they commute with each other.
They share the same $\mathbb{Z}_2$ subgroup, $R_V(\pi)=R_A(\pi)$, so globally the symmetry group is $[\mathrm{U(1)}_V\times \mathrm{U(1)}_A]/\mathbb{Z}_2$.

\subsection{The overlap Weyl fermion}
\label{sec:weyl}

L\"uscher imposes the following chiral projection on the Dirac fermion fields to define the overlap Weyl fermion model \cite{Luscher:1998du}:
\begin{equation}
\hP_-\psi=\psi,\qquad \bar\psi P_+=\bar\psi,\qquad S=\sum_{x,y}\bar\psi(x)D(x,y)\psi(y).
\label{eq:weyl}
\end{equation}
While the chiral projection of $\bar\psi$ is ultra-local, that of $\psi$ is local with exponential tails. 
The Dirac action splits as $\bar\psi D\psi=\bar\psi P_+D\hP_-\psi+\bar\psi P_-D\hP_+\psi$, and the Weyl theory keeps the first term. 
In the continuum limit, this lattice model flows to a single continuum Weyl fermion field theory.

At finite volume, the Grassmann measure is defined by choosing orthonormal bases $v_j$ of $\im\hP_-$ and $\bar v_k$ of $\im P_+$, expanding $\psi=\sum_jv_jc_j$, $\bar\psi=\sum_k\bar c_k\bar v_k^\dagger$, and integrating over the coefficients; a change of basis multiplies the measure by a phase, which is L\"uscher's phase ambiguity \cite{Luscher:1998pqa,Luscher:1998du,Luscher:1999un,Luscher:2000zd}.

After the projection only one $\mathrm{U(1)}$ global symmetry survives, which is\footnote{The axial rotation acts as $e^{i\varphi\hg}\psi=e^{-i\varphi}\psi$ and $\bar\psi e^{i\varphi\g_5}=e^{i\varphi}\bar\psi$, which is just a vector rotation with the opposite angle in the projected field space.}   
\begin{equation}
\mathrm{U(1)}:\quad \psi\to e^{i\varphi}\psi, \quad\bar\psi\to\bar\psi e^{-i\varphi}.
\end{equation}
This becomes the $\mathrm{U(1)}$ of the continuum Weyl fermion, which is anomalous. 
For a left-handed Weyl fermion of charge one, the anomaly is described by the anomaly polynomials
\begin{equation}
I_4=\frac12\Big(\frac{F}{2\pi}\Big)^2-\frac{p_1}{24}\quad(d=2),\qquad
I_6=\frac16\Big(\frac{F}{2\pi}\Big)^3-\frac{p_1}{24}\,\frac{F}{2\pi}\quad(d=4),
\label{eq:anompoly}
\end{equation}
where $F$ is the field strength of a background $\mathrm{U(1)}$ gauge field and $p_1$ is the first Pontryagin class. The terms with $p_1$ describe the gravitational anomaly in $d=2$, captured by the chiral central charge $c_L-c_R=1$, and the mixed $\mathrm{U(1)}$--gravitational anomaly in $d=4$. 
The first terms in both cases capture the 't Hooft anomaly (i.e., obstruction to gauging) of the $\mathrm{U(1)}$ global symmetry. 

It is unusual that this anomalous $\mathrm{U(1)}$ acts ``on-site'' in the lattice model, in the sense that it sends a field at site $x$ to another one at the same site.  
An anomalous symmetry usually cannot be realized on-site in a lattice model with a local action and independent local degrees of freedom. 
This is because such on-site symmetries can be coupled to dynamical gauge fields in a straightforward way and the symmetry is therefore gaugeable and free of anomalies. 
Here this is possible because the Grassmann variables are not independent. 
The field space $\im\hP_-$ is defined by a chiral projection that relates fields from different sites, so it is not a product of independent on-site  fields. 
The anomaly reappears in the Grassmann measure on $\im\hP_-$: in the presence of background gauge fields its phase is not  gauge-invariant unless the anomaly cancels \cite{Luscher:1998du,Luscher:1999un}. Similar anomalous global symmetries have also been realized in Euclidean bosonic modified Villain lattice models \cite{Sulejmanpasic:2019ytl,Gorantla:2021svj,Berkowitz:2023pnz,Berkowitz:2026kud}.

\subsection{The Weyl bundle}
\label{sec:bundle}

We now take a closer look at the chiral projection \eqref{eq:weyl}. At each momentum $p\in T^d$, the Weyl field $\psi(p)$ takes values in the $-1$ eigenspace of $\hg(p)$, and as $p$ ranges over the Brillouin torus these eigenspaces form a vector bundle, which we call the \textit{Weyl bundle}.  Its topology will be crucial for our proof in Section~\ref{sec:notheta} that the overlap Weyl fermion is not reflection positive. 
The topological properties of these bundles were discussed in \cite{Neuberger:1998xn,JahnPawlowski2002Chiral}.

In $d=2$, we have
\begin{equation}
\hg(p)=\hat n(p)\cdot\vec\sigma,\qquad \hat n=(-\beta_2,\ \beta_1,\ \alpha)=\frac{1}{\omega}\big(-\sin p_2,\ \sin p_1,\ \cos p_1+\cos p_2-1\big).
\label{eq:P2d}
\end{equation}
Let $\mathcal L_p=\im\hP_-(p)$ be the one-dimensional vector space of $\psi(p)$ satisfying $\hat\gamma_5 (p ) \psi(p) = -\psi(p)$. The family of lines $\mathcal L_p$ is a smooth complex line bundle $\mathcal L$ over $T^2$, the Weyl bundle, and a Weyl field $\psi(p)\in\mathcal L_p$ is a section of it. 

We can view $\hat\gamma_5(p)$ as the Hamiltonian of a two-level quantum system, a qubit in a unit magnetic field $\hat n(p)$, with the momentum $p\in T^2$ playing the role of the parameters of the Hamiltonian. 
The Weyl field at momentum $p$ is then a ground state of this Hamiltonian, $\mathcal L$ is the bundle of ground states over the parameter space, and the connection below is the Berry connection \cite{Berry:1984jv}. The map $\hat n:T^2\to S^2$ sends $p=0$ to the north pole and the three would-be doublers $(\pi,0)$, $(0,\pi)$, $(\pi,\pi)$ to the south pole.

The standard local sections $u(p)\in \mathcal{L}_p$ of this bundle are
\begin{equation}
u_\text{N}(p)=\frac{1}{\sqrt{2(1+\hat n_3)}}\begin{pmatrix}-(\hat n_1-i\hat n_2)\\ 1+\hat n_3\end{pmatrix},\qquad
u_\text{S}(p)=\frac{1}{\sqrt{2(1-\hat n_3)}}\begin{pmatrix}-(1-\hat n_3)\\ \hat n_1+i\hat n_2\end{pmatrix},
\label{eq:charts}
\end{equation}
which satisfy $\hat\gamma_5(p)u(p)= - u(p)$. 
The first is well-defined wherever $\hat n$ is away from the south pole $\hat n_3=-1$, i.e., away from the would-be doublers. 
Near the south pole $\hat n_1-i\hat n_2\to0$ with a phase that depends on the direction of approach, so $u_\text{N}$ has no limit there. 
The second is well-defined away from the north pole $\hat n_3=1$, i.e., away from $p=0$. 
In the region where both are defined, they differ by a phase,
\begin{equation}
u_\text{S}(p)=\lambda(p)\,u_\text{N}(p),\qquad \lambda=\frac{\hat n_1+i\hat n_2}{\sqrt{\hat n_1^2+\hat n_2^2}}=e^{i\phi_{\hat n}},
\label{eq:lambda}
\end{equation}
where $\phi_{\hat n}$ is the azimuthal angle of $\hat n$. This is precisely the Wu--Yang description of the Dirac monopole \cite{Wu:1975es}.

The connection of $\mathcal L$ in the local section $u(p)$ is $a_\mu=-iu^\dagger\partial_{p_\mu}u$, with curvature $f=\partial_1a_2-\partial_2a_1=-i\tr\big(\hP_-[\partial_1\hP_-,\partial_2\hP_-]\big)$, where we have used $\hP_-=uu^\dagger$. 
With $\partial_\mu\hP_-=-\frac12\partial_\mu\hat n\cdot\vec\sigma$ and $[\partial_1\hat n\cdot\vec\sigma,\partial_2\hat n\cdot\vec\sigma]=2i(\partial_1\hat n\times\partial_2\hat n)\cdot\vec\sigma$, one finds $f=-\frac12\,\hat n\cdot(\partial_1\hat n\times\partial_2\hat n)$. The first Chern number is
\begin{equation}
C_1=\frac1{2\pi}\int_{T^2}f\,d^2p
=-\frac1{4\pi}\int_{T^2}\hat n\cdot(\partial_1\hat n\times\partial_2\hat n)\,d^2p
=-\deg\hat n
=-1 .
\label{eq:c1}
\end{equation}
Here $\deg\hat n$ is the number of times $\hat n:T^2\to S^2$ covers the sphere. It is counted by the preimages of the north pole. 
It is easy to see that 
$\hat n_3=1$ holds only at $p=0$. 
The sign of the degree can be easily fixed by expanding around $p=0$, from which we find $\deg\hat n=1$.

In $d=4$, 
\begin{equation}
\begin{aligned}
&\hg(p)= - \hat h_A \Gamma_A 
 = -\begin{pmatrix}\hat h_5&\hat h_k\sigma_k+i\hat h_4\\ \hat h_k\sigma_k-i\hat h_4&-\hat h_5\end{pmatrix},
\\
&\hat h=(\beta_1,\beta_2,\beta_3,\beta_4,-\alpha),\quad
\Gamma=(i\g_5\g_1,\dots,i\g_5\g_4,\g_5).
\end{aligned}
\label{eq:P4d}
\end{equation}
Here and below, $A=1,\ldots,5$ labels the five components
of the unit vector $\widehat h\in S^4$ embedded in $\mathbb{R}^5$, while $\mu=1,\ldots,4$ labels the
four spacetime directions, and $k=1,2,3$. 
We can view this as the Hamiltonian of a four-level quantum system whose two energy levels $\pm1$ are each doubly degenerate. 
The Weyl fields at momentum $p$ span its two-dimensional ground space $\mathcal L_p=\im\hP_-(p)$, and $\mathcal L$ is a rank-two complex vector bundle over $T^4$, the pullback by $\hat h:T^4\to S^4$ of the ground-state bundle over $S^4$. The latter is Yang's $\mathrm{SU(2)}$ generalization of the Dirac monopole \cite{Yang:1977qv}, which on $S^4$ is the BPST instanton \cite{Belavin:1975fg}, as we verify below.

An explicit local orthonormal frame of $\mathcal L$ is the $4\times2$ matrix
\begin{equation}
U(p)=\frac1{\sqrt{2(1+\hat h_5)}}\begin{pmatrix}(1+\hat h_5)\,1\\ \hat h_k\sigma_k-i\hat h_4\end{pmatrix},
\label{eq:U4d}
\end{equation}
as one checks from $\hg U=-U$ and $U^\dagger U=1$. It is well-defined wherever $\hat h_5\neq-1$, which is everywhere except $p=0$. Its connection $a_\mu=-iU^\dagger\partial_{p_\mu}U$ is a $2\times2$ hermitian matrix, with curvature $f_{\mu\nu}=\partial_\mu a_\nu-\partial_\nu a_\mu+i[a_\mu,a_\nu]$. 

Using the stereographic coordinates $x_\mu=\hat h_\mu/(1+\hat h_5)$ of $S^4$, we can rewrite the local frame and the connection as
\begin{equation}
U=\frac1{\sqrt{1+|x|^2}}\begin{pmatrix}1\\X\end{pmatrix},\qquad a=-\frac i2\,\frac{X^\dagger dX-dX^\dagger X}{1+|x|^2},
\end{equation}
where $X=x_k\sigma_k-ix_4$ and $X^\dagger X=|x|^2=x_\mu x_\mu$. 
This is precisely the unit BPST instanton \cite{Belavin:1975fg} in regular gauge, pulled back to the Brillouin zone by $p\mapsto x(p)$.

The integral of the second Chern character is therefore the instanton number times the degree of $\hat h:T^4\to S^4$. Explicitly,
\begin{equation}
C_2=\frac1{8\pi^2}\int_{T^4}\tr f\wedge f
=-\frac3{8\pi^2}\int_{T^4}\det\big(\hat h,\partial_1\hat h,\ldots,\partial_4\hat h\big)\,d^4p
=-\deg\hat h
=1 .
\label{eq:c2}
\end{equation}
The degree is counted by the preimages of the south pole $\hat h=(0,0,0,0,-1)$, which consist only of $p=0$.
Near the south pole, $\hat h\simeq(p_1,p_2,p_3,p_4,-1)$ and $\det(\hat h,\partial_1\hat h,\ldots,\partial_4\hat h)=-1$, so $\hat h$ reverses orientation there and $\deg\hat h=-1$. 
The topological invariants $C_1$ and $C_2$ computed above coincide with the Chern--Simons levels of the $(d+1)$-dimensional Wilson fermion found in \cite{Golterman:1992ub}.

\section{Reflection positivity}
\label{sec:rp}

A reflection-positive Euclidean lattice theory admits
a reconstruction of a Hilbert space and a positive
transfer matrix, and hence a self-adjoint Hamiltonian 
\cite{Osterwalder:1973dx,Osterwalder:1974tc}.
For lattice gauge theories with Wilson fermions this was established in \cite{Osterwalder:1977pc,Luscher:1976ms,Menotti:1987pu}; see \cite{Zenkin:1992ax,Zenkin:1992ts} for earlier attempts at reflection-positive chiral lattice theories and \cite{Montvay:1994cy} for a review.

Observables are polynomials in the fields and $\langle F\rangle$ denotes their normalized expectation value. 
Throughout this section we work on the infinite lattice $\mathbb Z^d$, with $x_d$ the Euclidean time direction.\footnote{On a finite torus one imposes antiperiodic boundary conditions in time and restricts $\mathcal A_\pm$ below to the corresponding halves of the torus.} 
Let $\theta(\vec x,x_d)=(\vec x,1-x_d)$ be the reflection about the plane between the slices $x_d=0$ and $x_d=1$. 
Let $\mathcal A_+$ be the polynomials in the fields with $x_d\ge1$, and $\mathcal A_-$ those with $x_d\le0$. 
A reflection is an antilinear involution $\Theta$ of the Grassmann algebra that reverses the order of products, $\Theta(FG)=\Theta(G)\Theta(F)$, and exchanges $\mathcal A_+$ and $\mathcal A_-$. The theory is reflection positive if $\Theta(S)=S$ and
\begin{equation}
\langle\Theta(F)\,F\rangle\ge0\qquad\text{for all }F\in\mathcal A_+ .
\label{eq:rp}
\end{equation}
The Hilbert space is then $\mathcal A_+$ with the inner product $\langle\Theta(F)G\rangle$ modulo null vectors.

\subsection{Axial symmetry and reflection positivity}

We start with the overlap Dirac fermion model, which is reflection positive \cite{Kikukawa:2010gq} with respect to the reflection \cite{Osterwalder:1977pc,Montvay:1994cy}
\begin{equation}
\TD\psi(x)=\bar\psi(\theta x)\,\g_d,\qquad \TD\bar\psi(x)=\g_d\,\psi(\theta x).
\label{eq:theta}
\end{equation}
Let us check the invariance of the action. Consider a quadratic action $\sum_{x,y}\bar\psi(x)K(x,y)\psi(y)$, where $K(x,y)$ is a $2^{d/2}\times2^{d/2}$ matrix for each pair of sites. Since $\TD$ is antilinear and reverses the order of products,
\begin{equation}
\TD\Big(\sum_{x,y}\bar\psi(x)K(x,y)\psi(y)\Big)=\sum_{x,y}\bar\psi(\theta y)\,\g_dK(x,y)^\dagger\g_d\,\psi(\theta x)=\sum_{x,y}\bar\psi(x)\,\g_dK(\theta y,\theta x)^\dagger\g_d\,\psi(y),
\end{equation}
where $K(x,y)^\dagger$ is the hermitian conjugate of the matrix $K(x,y)$. It is convenient to define the reflected kernel $K^{\TD}(x,y)=\g_dK(\theta x,\theta y)\g_d$, and the adjoint kernel $K^\dagger(x,y)=K(y,x)^\dagger$. Then $\g_dK(\theta y,\theta x)^\dagger\g_d=(K^{\TD})^\dagger(x,y)$, and the quadratic action is invariant if and only if
$K^{\TD}(x,y)=K^\dagger(x,y)$. 
It is easy to see that the Wilson operator satisfies $D_{\rm w}^{\TD}=D_{\rm w}^\dagger$.
Since the overlap operator is a real function of $D_{\rm w}$, it follows that 
\begin{equation}\label{eq:Dtheta}
D^{\TD}=D^\dagger=\g_5D\g_5 ,\quad \g_5^{\TD}=-\g_5.
\end{equation}
Hence both the Wilson and the overlap Dirac actions are invariant.

The two global symmetries of the overlap Dirac fermion behave differently under $\TD$. Denote by $R_V(\varphi)$ and $R_A(\varphi)$ the transformations \eqref{eq:V} and \eqref{eq:A} acting on the Grassmann algebra. The vector symmetry is invariant under the reflection, 
\begin{equation}
\TD\,R_V(\varphi)\,\TD=R_V(\varphi).
\end{equation}
Furthermore, $\mathrm{U(1)}_V$ maps the positive-time algebra $\mathcal A_+$ to itself. 
It therefore leads to a unitary operator acting on the Hilbert space reconstructed from $\TD$, and the vector charge is a self-adjoint operator with integer eigenvalues.  See Appendix~\ref{app:unitary}. 

On the other hand, the axial symmetry is not invariant. From $\hg=\g_5(1-D)$ and \eqref{eq:Dtheta},
\begin{equation}
\hg^{\TD}
=-(1-D)\,\g_5 ,
\label{eq:hgtheta}
\end{equation}
and therefore 
\begin{equation}
\TD\,R_A(\varphi)\,\TD=\tilde R_A(\varphi),
\end{equation}
where $
\tilde R_A(\varphi)$ implements another $\widetilde{\mathrm{U(1)}}_A$ transformation\footnote{More generally, using \eqref{eq:theta}, the reflection $\TD$ maps a transformation $\psi\to A\psi$, $\bar\psi\to\bar\psi B$ to $\psi\to(B^{\TD})^\dagger\psi$, $\bar\psi\to\bar\psi(A^{\TD})^\dagger$. }
\begin{equation}
\tilde R_A(\varphi):\quad\psi\to e^{i\varphi\g_5}\psi,\qquad \bar\psi\to\bar\psi\,e^{i\varphi(1-D)\g_5}.
\label{eq:mirrorA}
\end{equation}
This transformation is also a symmetry of the action, but $\tilde R_A(\varphi)\neq R_A(\varphi)$ for $\varphi\neq0,\pi$. 
In fact, the two transformations $\mathrm{U(1)}_A$ and $\widetilde{\mathrm{U(1)}}_A$ generate an Onsager algebra as reviewed in Appendix~\ref{app:onsager}.

Even though $\mathrm{U(1)}_A$ and $\widetilde{\mathrm{U(1)}}_A$ are separately not invariant under the reflection, their average is. Averaging the generators of \eqref{eq:A} and \eqref{eq:mirrorA} gives the transformation:
\begin{equation}
R_{\rm av}(\varphi):\quad \psi\to e^{i\varphi\g_5(1-\frac12D)}\psi,\qquad \bar\psi\to\bar\psi\,e^{i\varphi(1-\frac12D)\g_5},
\label{eq:qA}
\end{equation}
which is a symmetry of the action. Since $\TD$ exchanges $\mathrm{U(1)}_A$ and $\widetilde{\mathrm{U(1)}}_A$, we have 
\begin{equation}
\TD\,R_{\rm av}(\varphi)\,\TD=R_{\rm av}(\varphi).
\end{equation}
The eigenvalues of $\gamma_5(1-\frac12 D)$ are
$\pm\sqrt{(1+\alpha(p))/2}$, which vary continuously
with momentum. Hence the transformation $R_{\rm av}(\varphi)$
is not periodic in $\varphi$ and generates a noncompact
group $\mathbb R$ rather than U(1).
This is the Euclidean counterpart of the conserved but unquantized axial charge $\g_5(1-\frac a2D_\text{s})$ of \cite{Creutz:2001wp,Singh:2025sye,Kourkoulou:2026lzx} in the Hamiltonian formulation of overlap fermions, where $D_\text{s}$ is the spatial Ginsparg--Wilson operator. 

To summarize, the overlap Dirac fermion has a compact axial symmetry, $\mathrm{U(1)}_A$, that is not invariant under the reflection, and a reflection-invariant axial symmetry, \eqref{eq:qA}, that is not compact.

 \subsection{The overlap Weyl fermion is not reflection positive}
\label{sec:notheta}

We now turn to the overlap Weyl fermion.\footnote{We thank Aleksey Cherman and Yoshio Kikukawa for insightful discussions on various points in this subsection.} 
The immediate difficulty is that the chiral projection of $\psi$ in \eqref{eq:weyl},
\begin{equation}
\hP_-\psi=\psi\quad\Leftrightarrow\quad \sum_y\hg(x,y)\psi(y)=-\psi(x),
\end{equation}
relates the fields at all times. It is therefore not clear how to define the positive-time algebra ${\cal A}_+$. 

The authors of \cite{Kikukawa:2010gq} address this by restricting to the observables built from $q(x)=P_-\psi(x)$ and $\bar q(x)=\bar\psi(x)P_+$, and they prove reflection positivity for the algebra generated by these fields at positive times. 
This is natural from the point of view of domain-wall fermions: in the limit of an infinite extra dimension, the correlation functions of $q,\bar q$ coincide with those of the boundary chiral modes of the domain-wall fermion \cite{Vranas:1997da,Neuberger:1997bg,Kikukawa:2000ac}, which is a Wilson fermion in one higher dimension and is reflection positive. 
The path integral is still defined by the overlap action \eqref{eq:weyl}. 
This set of observables, however, does not include all the fields of the model: the component $P_+\psi(x)$ is not generated by $q,\bar q$ at positive times. 
Indeed, using $\hP_+\psi=0$, one finds $\psi=\frac{2}{1+\alpha}\hP_-P_-q$ in momentum space, which is nonlocal in position space and singular at the would-be doublers (i.e., the zeros of $\sin p_\mu$ other than $p=0$), where $\alpha=-1$ and $q$ vanishes.\footnote{The correlation functions of $q,\bar q$ are those of a free theory with the kernel \cite{Chiu:1998gp,Chiu:1998eu,Edwards:1998wx,Kikukawa:2000ac,Edwards:2000qv}
\begin{equation}\label{eq:Dq}
D_q(p)=\frac{2i}{1+\alpha(p)}\,P_+\g_\mu\beta_\mu(p)P_-\,,
\end{equation}
which diverges at the would-be doublers. This singularity violates the assumption on the kernel in the Euclidean no-go theorems \cite{Karsten:1981gd,Karsten:1980wd,Baig:2026zgb}.}
Thus, reflection positivity holds for the observables built from $q,\bar q$, but, as we discuss below, not for the full algebra of Weyl fields.

A related problem is that $\TD$, extended to all fields by \eqref{eq:theta}, does not preserve the chiral projection \eqref{eq:weyl}. 
Using \eqref{eq:hgtheta}, $\TD$ maps \eqref{eq:weyl} to
\begin{equation}
P_-\psi=\psi,\qquad \bar\psi\,\tfrac12\big(1+(1-D)\g_5\big)=\bar\psi ,
\end{equation}
which defines a different field space. Therefore $\TD$ cannot be used to formulate reflection positivity for the overlap Weyl fermion.

 One may ask whether the chiral projection \eqref{eq:weyl} of the  overlap Weyl fermion can be preserved by some other antilinear map $\Theta$ that respects locality and is compatible with the symmetries. The answer is no, and the reason is topological. Below we show that, on an infinite lattice $\mathbb{Z}^d$, the existence of such a reflection implies a nowhere-vanishing global section of the Weyl bundle of Section~\ref{sec:bundle}. However, the Weyl bundles have nonzero top Chern classes, so no such section exists. Hence, the overlap Weyl fermion is not reflection positive with respect to any such  reflection.

The overlap Weyl fermion is the Dirac Grassmann algebra subject to the chiral projection \eqref{eq:weyl}, i.e., to the linear relations $\hP_+\psi=0$ and $\bar\psi P_-=0$. We ask for a map $\Theta$ that satisfies the following conditions:
\begin{enumerate}\setlength{\itemsep}{0pt}
\item It is an antilinear map of the Weyl fields with $\Theta^2=1$; in particular it preserves the chiral projection \eqref{eq:weyl}, mapping the relations $\hP_+\psi=0$ and $\bar\psi P_-=0$ to combinations of themselves;
\item It commutes with the $\mathrm{U(1)}$ symmetry, so that the charge is self-adjoint on the reconstructed Hilbert space; since $\Theta$ is antilinear, it must then map $\psi$ to $\bar\psi$ and $\bar\psi$ to $\psi$;
\item It is covariant under translations composed with the reflection $\theta$;
 \item It is local with exponential tails: $\Theta\psi(x)$  is a combination of the fields $\bar\psi(y)$ whose coefficients decay exponentially in $|y-\theta x|$. A similar condition holds for $\Theta\bar\psi(x)$. 
\end{enumerate}
(These conditions do not require $\Theta$ to map $\mathcal A_+$ into $\mathcal A_-$ exactly, so the result below rules out a larger class of maps than reflections in the sense above.)
Conditions 2 and 3 fix the form of $\Theta$ on $\psi$:
\begin{equation}
\Theta\psi(x)=\sum_y\bar\psi(y)\,A(y-\theta x),
\label{eq:Thetagen}
\end{equation}
with a $2^{d/2}\times 2^{d/2}$ matrix kernel $A$; the Dirac reflection \eqref{eq:theta} corresponds to $A(z)=\delta_{z,0}\g_d$. Condition 4 says that $A(z)$ decays exponentially in $|z|$, so that its Fourier transform $\tilde A(q)=\sum_ze^{-iq\cdot z}A(z)$ is analytic and periodic. 

In momentum space, with  $\theta p=(\vec p,-p_d)$, \eqref{eq:Thetagen} becomes\footnote{The phase $e^{ip_d}$ comes from the one-unit shift in the link reflection $\theta x=(\vec x,1-x_d)$ and is absent for the site reflection $(\vec x,x_d)\to(\vec x,-x_d)$. The argument below does not depend on this phase, so our conclusion also holds for the site reflection.}
\begin{equation}
\Theta\psi_a(p)=e^{ip_d}\sum_b\bar\psi_b(\theta p)\,\tilde A_{ba}(\theta p).
\label{eq:Thetap}
\end{equation}
Condition 1 demands that the chiral projection be preserved. Applying $\Theta$ to $\hP_+(p)\psi(p)=0$ and using \eqref{eq:Thetap} and antilinearity gives $\bar\psi(\theta p)\tilde A(\theta p)\hP_+(p)^\dagger=\bar\psi(\theta p)\tilde A(\theta p)\hP_+(p)=0$, where we used $\hP_+(p)^\dagger=\hP_+(p)$. The only linear relations among the components of $\bar\psi$ are $\bar\psi P_-=0$, so we must have
\begin{equation}
P_+\tilde A(\theta p)\,\hP_+(p)=0,\qquad\forall p .
\label{eq:constraintmap}
\end{equation}

In $d=2$, $\bar\psi P_-=0$ implies that the Weyl theory has a single component $\bar\psi_1$ of $\bar\psi$ at each momentum, and only the first row of $\tilde A(\theta p)$ enters \eqref{eq:Thetap}. Calling this row $r(p)$,
\begin{equation}
P_+\tilde A(\theta p)=\begin{pmatrix}r(p)\\0\end{pmatrix},\qquad \Theta\psi(p)=e^{ip_d}\bar\psi_1(\theta p)\,r(p)^T.
\label{eq:Thetarow}
\end{equation}
Condition \eqref{eq:constraintmap} says $r(p)\hP_+(p)=0$. In other words, the vector $r(p)^\dagger$ lies in $\mathcal L_p=\im\hP_-(p)$, the line of allowed Weyl spinors at momentum $p$. 
Since $\Theta$ is invertible, $r(p)\neq0$. 
By condition 4, $r(p)$ is analytic and periodic. Thus $r(p)^\dagger$ is a nonzero vector in $\mathcal L_p$ for every $p$, varying smoothly over the Brillouin torus: a global section of the Weyl bundle $\mathcal L$ of Section~\ref{sec:bundle} that vanishes nowhere. 
However, since $C_1=-1$ in \eqref{eq:c1}, no such section exists. 
So no $\Theta$ satisfying conditions 1--4 exists in $d=2$.

In $d=4$ the argument is the same.  Now $P_+=\mathrm{diag}(1,1,0,0)$, the Weyl theory has two components $\bar\psi_1,\bar\psi_2$ of $\bar\psi$ at each momentum, and only the first two rows of $\tilde A(\theta p)$ enter \eqref{eq:Thetap}. Calling them $r_1(p),r_2(p)$, we have
\begin{equation}
\begin{gathered}
P_+\tilde A(\theta p)=\begin{pmatrix}r_1(p)\\ r_2(p)\\ 0\\ 0\end{pmatrix},\\
\Theta\psi(p)=e^{ip_d}\Big[\bar\psi_1(\theta p)\,r_1(p)^T+\bar\psi_2(\theta p)\,r_2(p)^T\Big] .
\end{gathered}
\label{eq:Thetarow4}
\end{equation}
Equation \eqref{eq:constraintmap} implies  $r_i(p)^\dagger\in\mathcal L_p$, which is now a two-dimensional vector space. 
Since $\Theta$ is invertible,  $r_1(p)\neq0$ at every $p$. Indeed, if $r_1(p)$ were zero for some $p$, \eqref{eq:Thetarow4} would map the two independent components of $\psi(p)\in\mathcal L_p$ into the one-dimensional span of $\bar\psi_2(\theta p)$, so some nonzero combination of them would be mapped to zero. 
Thus, $r_1(p)^\dagger$ is a nowhere-vanishing global section of the rank-two bundle $\mathcal L$ over $T^4$. 
But this is a contradiction: for a rank-two bundle over a four-dimensional base, the second Chern class obstructs such a section, and it is nonzero for $\mathcal L$ as discussed in around \eqref{eq:c2}.\footnote{For the $d=4$ Weyl bundle, the second Chern class equals minus the second Chern character, because the first Chern class vanishes with the structure group being $\mathrm{SU(2)}$. }  
So no $\Theta$ satisfying conditions 1--4 exists in $d=4$. 
The result of \cite{Kikukawa:2010gq} does not contradict this, since it concerns a restricted set of observables.

In $d=4$, we can rule out such an antilinear reflection even without assuming condition 2. 
Without the $\mathrm{U(1)}$ symmetry, $\Theta\psi(x)$ can be a combination of both $\bar\psi(y)$ and $\psi(y)$ with exponentially decaying coefficients, and similarly for $\Theta\bar\psi(x)$. 
In momentum space, \eqref{eq:Thetap} is replaced by
\begin{equation}
\Theta\psi_a(p)=e^{ip_d}\sum_b\Big[\bar\psi_b(\theta p)\,\tilde A_{ba}(\theta p)+\tilde B_{ab}(\theta p)\,\psi_b(-\theta p)\Big].
\label{eq:ThetapnoU1}
\end{equation}
Similarly, $\Theta\bar\psi(-p)$ is a combination of $\bar\psi(\theta p)$ and $\psi(-\theta p)$, with coefficients that are analytic and periodic in $p$. 
Define $\Phi(p)=\big(\psi(p),\bar\psi(-p)^T\big)^T$.  We have
\begin{equation}
\Theta\Phi(p)=N(p)\,\Phi(-\theta p),
\label{eq:ThetaPhi}
\end{equation}
with an $8\times8$ matrix $N(p)$ that is analytic and periodic. 
The chiral projection \eqref{eq:weyl} now reads
\begin{equation}
\Pi(p)\,\Phi(p)=\Phi(p),\qquad \Pi(p)=\begin{pmatrix}\hP_-(p)&0\\0&P_+\end{pmatrix},
\end{equation}
where $\Pi(p)$ is a projector of rank four. Its image is $\mathcal L_p\oplus\im P_+$, so the bundle $\im\Pi$ over $T^4$ has the same second Chern number as the Weyl bundle, i.e., $C_2=1$. 
As in the derivation of \eqref{eq:constraintmap}, condition 1 requires
\begin{equation}
\big(1-\Pi(p)^*\big)\,N(p)\,\Pi(-\theta p)=0,\qquad\forall p .
\end{equation}
Hence $N(p)$ maps $\im\Pi(-\theta p)$ to $\im\Pi(p)^*$, and this map is invertible because $\Theta$ is. 
The two bundles over $T^4$ with these fibers are therefore isomorphic and have the same Chern numbers. 
The second is the complex conjugate of the bundle $\im\Pi$, and the first is its pullback by the orientation-reversing map $p\mapsto-\theta p=(-\vec p,p_4)$ of $T^4$. 
For a bundle over $T^{2k}$, complex conjugation changes the Chern number $C_k$ to $(-1)^kC_k$, while the orientation reversal changes $C_k$ to $-C_k$. 
In $d=4$ this gives $C_2=-C_2$, which contradicts $C_2=1$. 
So no $\Theta$ satisfying conditions 1, 3, and 4 exists in $d=4$. 
In $d=2$, by contrast, both operations reverse the sign of $C_1$ and there is no contradiction.

We emphasize that the obstruction to reflection positivity at finite
lattice spacing need not persist in the continuum limit.
Indeed, the chiral projection implies
$P_+\psi=\frac12 P_+D\psi$, which is of order $a$ in the
continuum limit, since $D(p)\sim ia\gamma_\mu k_\mu$
for fixed physical momentum $k_\mu=p_\mu/a$.
Thus, the component $P_+\psi$ that obstructs the Dirac reflection
$\TD$ vanishes in the continuum limit.
Furthermore, our topological obstruction relies on the
nontriviality of the Weyl bundle over the entire Brillouin torus,
whereas the continuum limit probes only a neighborhood of
$p=0$, over which the bundle is trivial.
We therefore expect reflection positivity to be recovered
in the continuum limit, consistent with the unitarity of the
continuum Weyl fermion.

\section{Uncuttable lattice and gravitational anomalies}
\label{sec:bdy}

So far we have worked on a closed lattice. In this section we cut the lattice open and ask whether the overlap Weyl fermion remains local. We restrict to $d=2$ for simplicity.

We work on the half-plane, whose sites are $x=(x_1,x_2)$ with $x_1=0,1,2,\ldots$ and $x_2\in\mathbb Z$, so that the boundary is the column $x_1=0$.\footnote{In the numerics we use a strip $0\le x_1\le L_1-1$ with $x_2\sim x_2+L_2$. For $L_1,L_2$ large, the results below hold up to finite-size corrections near the boundary $x_1=0$.} 
Naively, one might truncate the overlap operator $D(x,x')$ of the closed lattice to the sites with $x_1,x_1'\ge0$. This does not work. The chiral projection \eqref{eq:weyl} used in the overlap Weyl fermion requires the modified chirality to satisfy $\hg^2=1$, so that $\hP_\pm$ are projectors. This follows from the Ginsparg--Wilson relation \eqref{eq:GW}, which the naive truncation of $D(x,x')$ violates near the boundary.

For the overlap Dirac fermion, L\"uscher constructed a different boundary that keeps $D$ exponentially local, at the price of violating the Ginsparg--Wilson relation, and hence the axial symmetry, near the boundary \cite{Luscher:2006df}. This is not an option for the overlap Weyl fermion, since the chiral projection \eqref{eq:weyl} cannot be imposed without the Ginsparg--Wilson relation.

Instead, we truncate the ultra-local Wilson operator $D_{\rm w}$ and construct $D$ from it.\footnote{More precisely, $D_{\rm w}$ on the half-plane is given by the same expression as on the closed lattice, except that the hopping terms between the columns $x_1=0$ and $x_1=-1$ are removed. Equivalently, we impose the Dirichlet boundary condition $\psi(-1,x_2)=0$.} The resulting $H_{\rm w}=\g_5(D_{\rm w}-1)$ is still hermitian. As long as it has no zero eigenvalue, $D=1+\g_5\,\mathrm{sgn}(H_{\rm w})$ is well-defined and satisfies the Ginsparg--Wilson relation, and the chiral projection \eqref{eq:weyl} can be imposed as before. The question is whether $D$ is still local.

There is a reason to expect that it is not. The continuum Weyl fermion has a gravitational anomaly captured by the chiral central charge $c_L- c_R=1$, and a two-dimensional theory with a gravitational anomaly admits no boundary condition \cite{Cardy:1989ir,Jensen:2017eof,Thorngren:2020yht,Hellerman:2021fla}. Physically, a boundary must stop the flow of energy across it, so it must reflect every incoming mode into an outgoing one. A single Weyl fermion moves in only one direction and has nothing to reflect into.
Below we show that the nonlocality appears already in the overlap operator $D$. On a closed lattice $|H_{\rm w}|\ge1$, which is what makes $D$ exponentially local. On the half-plane, $H_{\rm w}$ acquires a zero eigenvalue at zero momentum along the boundary, and this makes $D$ decay only as a power along the boundary.

The half-plane is translation invariant along the boundary, so we Fourier transform in $x_2$, with momentum $k$, while keeping $x_1$ in position space,
\begin{equation}
D(x,x')=\int_{-\pi}^{\pi}\frac{dk}{2\pi}\,e^{ik(x_2-x_2')}\,D(k)_{x_1x_1'} .
\end{equation}
Then $H_{\rm w}$ becomes an operator $H_{\rm w}(k)$ acting on a spinor $\Psi(x_1)$ as
\begin{equation}
\begin{aligned}
&H_{\rm w}(k)\,\Psi(x_1)=\sigma_1\sin k\,\Psi(x_1)+\sigma_3\Big[\lambda_k\,\Psi(x_1)-\tfrac12(1-\sigma_1)\,\Psi(x_1+1)-\tfrac12(1+\sigma_1)\,\Psi(x_1-1)\Big],\\
&\lambda_k=1-\cos k,
\label{eq:H2d}
\end{aligned}
\end{equation}
where the last term is absent at $x_1=0$. 
If $D(k)_{x_1x_1'}$ is analytic in $k$, $D(x,x')$ decays exponentially in $x_2-x_2'$, while a singularity produces a power law. 
Such a singularity can only occur at momenta where $H_{\rm w}(k)$ has a zero eigenvalue.

The two terms in \eqref{eq:H2d} anticommute, so $H_{\rm w}(k)^2\ge\sin^2k$, and at $k=\pi$ the second term alone gives $H_{\rm w}(\pi)^2\ge1$. Hence a zero eigenvalue can occur only at $k=0$. For $\lambda_k<1$, i.e., $|k|<\pi/2$, $H_{\rm w}(k)$ has an exact eigenvector localized at the boundary,
\begin{equation}
H_{\rm w}(k)\,\Psi_k=-\sin k\;\Psi_k,\qquad \Psi_k(x_1)=\sqrt{1-\lambda_k^2}\;\lambda_k^{\,x_1}\Psi_0,
\label{eq:edge}
\end{equation}
where $\Psi_0$ is a fixed spinor obeying $\sigma_1\Psi_0=-\Psi_0$.
See \cite{Taniguchi:2004gf} for a related discussion. 
This leads to a factor of $\mathrm{sgn}(k)$ in the overlap operator $D(k)_{x_1x_1'}$ through $D=1+\sigma_3 \,\mathrm{sgn}(H_\text{w})$. 
Using $\int\frac{dk}{2\pi}e^{ikr}\,\mathrm{sgn}(k)\simeq\frac{i}{\pi r}$ at large $r$, we obtain
\begin{equation}
D\big((0,x_2),(0,x_2')\big)\simeq 
{Y\over x_2-x_2'} ,
\label{eq:tail}
\end{equation}
with some constant matrix $Y$.\footnote{A more detailed calculation gives $Y=-\frac{i}{\pi}\,\sigma_3\,\frac{1-\sigma_1}{2}$.}

\begin{figure}[h!]
\centering
\includegraphics[width=1\textwidth]{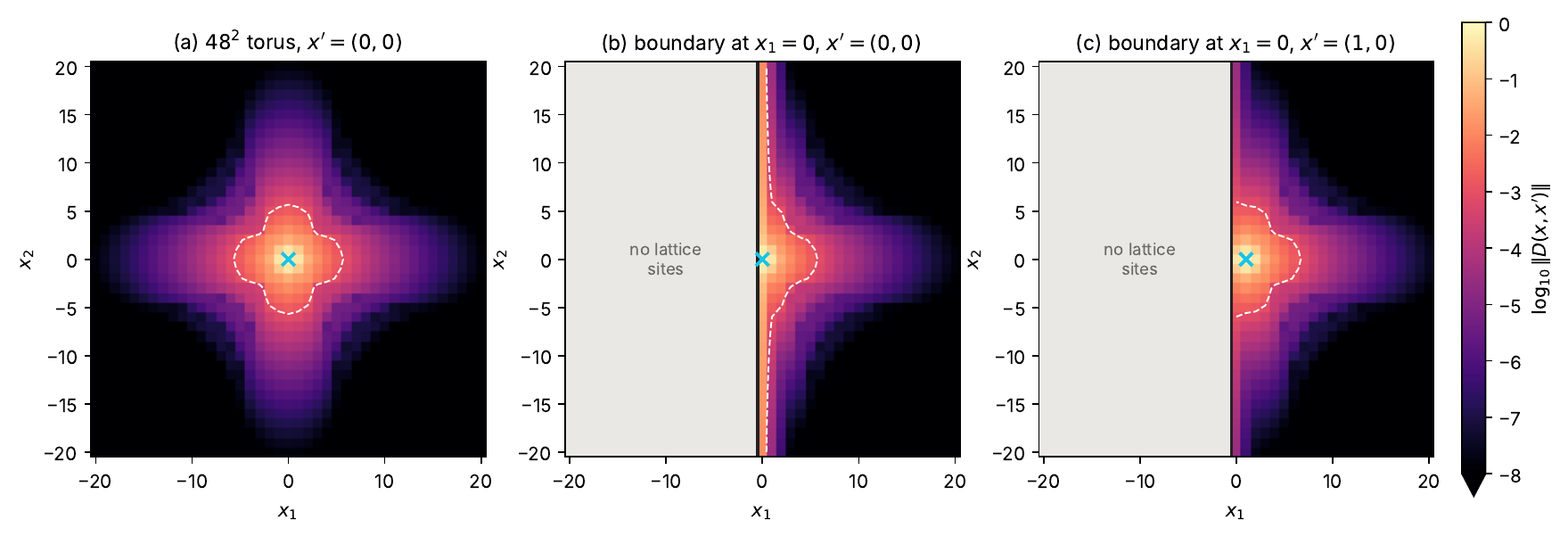}
\caption{$\log_{10}\|D(x,x')\|$ as a function of $x$ for fixed $x'$ (cross), with $\|M\|=(\frac12\tr M^\dagger M)^{1/2}$, from exact diagonalization of $H_{\rm w}$ in position space. (a)~Antiperiodic $48^2$ torus. (b)~Boundary at $x_1=0$ with $x'$ on the boundary, on a strip with $L_1=32$ and antiperiodic $L_2=96$. (c)~Same strip with $x'$ one site from the boundary. The dashed contour marks $\|D(x,x')\|=10^{-3}$. 
The decay is exponential in (a), and power-law in (b) and (c) along the boundary.
}
\label{fig:Dmaps}
\end{figure}

Figure~\ref{fig:Dmaps} confirms this power-law tail numerically. Without a boundary $D$ decays exponentially in every direction. With a boundary and $x'$ on the boundary, its weight spreads into a band along the entire boundary. 
The power law becomes steeper as $x'$ moves away from the boundary.

\section*{Acknowledgements}

We thank Tom Banks, Simon Catterall, Aleksey Cherman, Ting-Wai Chiu, Minjae Cho, Nathaniel Craig, Daniel Harlow, David B. Kaplan, Yoshio Kikukawa, Ioanna Kourkoulou, Sahand Seifnashri, Srimoyee Sen, Nikita Sopenko, David Tong, and Cenke Xu for discussions. 
We are especially grateful to Aleksey Cherman for stimulating discussions that inspired Section \ref{sec:notheta}, and for bringing \cite{Kikukawa:2010gq} to our attention. 
We also thank Yoshio Kikukawa for thoughtful comments on a draft. 
S.H.S. was supported in part by the U.S. Department of Energy, Office of Science, Office of High Energy
Physics under Contract Number DE-SC0012567, and in part by the Wade
Fund at the School of Science at MIT. 
The discussion and proof related to the Weyl bundle in $d=4$, some of the derivations in Section \ref{sec:bdy}, and the numerical plot were assisted
by GPT-6 Astra, Claude Opus 5.5, Fable 5.1, and Fable 5.2.
The authors of this paper are ordered alphabetically.

\appendix

\section{Global symmetries on the reconstructed Hilbert space}
\label{app:unitary}

In this appendix we briefly review how a symmetry transformation of a reflection-positive Euclidean lattice theory leads to a unitary operator acting on the reconstructed Hilbert space. See \cite{Neeb:2018book} for a systematic treatment. 

Let $\mathcal A_+$ be the algebra of observables supported at positive Euclidean times. 
Let $\Theta$ be the antilinear reflection.
We denote the Euclidean expectation value by $\langle\,\cdot\,\rangle$. Reflection positivity \eqref{eq:rp} states that the sesquilinear form
$B(F,G)=\big\langle\Theta(F)\,G\big\rangle$ with $ F,G\in\mathcal A_+$
is positive semidefinite. The reconstructed Hilbert space is obtained by dividing out the null vectors and completing \cite{Osterwalder:1973dx,Osterwalder:1974tc},
\begin{equation}
\mathcal H=\overline{\mathcal A_+/\mathcal N},\qquad \mathcal N=\{F\in\mathcal A_+:B(F,F)=0\},\qquad \big\langle[F],[G]\big\rangle_{\mathcal H}=B(F,G),
\end{equation}
where $[F]$ denotes the class of $F$.

Let $R_\varphi$ be a $\mathrm{U(1)}$ symmetry acting on the Grassmann algebra by linear automorphisms, with $R_{\varphi+2\pi}=R_\varphi$, which leaves all expectation values invariant, i.e., $\langle R_\varphi(X)\rangle=\langle X\rangle$. Assume in addition that it preserves the positive-time algebra, $R_\varphi\mathcal A_+=\mathcal A_+$. The candidate unitary operator $U_\varphi$ on the Hilbert space is
\begin{equation}
U_\varphi[F]=[R_\varphi F].
\end{equation}
For this to be well-defined and unitary, $U_\varphi$ must preserve the inner product. Using the invariance of the expectation value under $R_{-\varphi}$,
\begin{equation}
\big\langle U_\varphi[F],U_\varphi[G]\big\rangle_{\mathcal H}=\big\langle\Theta(R_\varphi F)\,R_\varphi G\big\rangle=\big\langle\big(R_{-\varphi}\Theta R_\varphi F\big)\,G\big\rangle .
\end{equation}
Therefore $U_\varphi$ preserves the inner product if and only if
\begin{equation}
\Big\langle\Big[\big(R_{-\varphi}\Theta R_\varphi-\Theta\big)F\Big]\,G\Big\rangle=0\qquad\text{for all }F,G\in\mathcal A_+\text{ and all }\varphi .
\label{eq:unitarycond}
\end{equation}
A sufficient condition for \eqref{eq:unitarycond} is that the symmetry is invariant under the reflection, $\Theta R_\varphi\Theta=R_\varphi$, which is the condition used in Section~\ref{sec:rp}; see Lemma 1.4 of \cite{Neeb:2012ts}. 
If \eqref{eq:unitarycond} holds, the map $\varphi\mapsto U_\varphi$ is a unitary representation of $\mathrm{U(1)}$, and by Stone's theorem $U_\varphi=e^{i\varphi Q}$ with a self-adjoint charge $Q$, whose eigenvalues are integers because $U_{2\pi}=1$.

\section{The Onsager algebra of the overlap Dirac fermion}
\label{app:onsager}

In Section~\ref{sec:rp} we discussed three $\mathrm{U(1)}$ symmetries of the overlap Dirac action $\bar\psi D\psi$: the vector symmetry U(1)$_V$ \eqref{eq:V}, L\"uscher's axial symmetry U(1)$_A$ \eqref{eq:A}, and its reflection image $\widetilde{\mathrm{U(1)}}_A$ \eqref{eq:mirrorA} under the Dirac reflection. Here we work out the symmetry they generate. The two axial symmetries generate an Onsager algebra \cite{Onsager:1943jn,Dolan:1981sh}. Onsager algebras have recently been found to act as exact symmetries of many free lattice fermions in the Hamiltonian formalism, where they emanate the chiral symmetry of the continuum \cite{Chatterjee:2024gje,Xu:2025hfs,Gioia:2025bhl,Pace:2025rfu,Catterall:2025vrx,Yamaoka:2025sdm,Onogi:2025xir,Aoki:2025vtp,Kim:2025tzc,Misumi:2026ckr,Lew-Smith:2026jqb,Kourkoulou:2026lzx} (see also \cite{Vernier:2018han,Pace:2024oys}). In the context of the Ginsparg--Wilson relation, the symmetries below were found by Mandula \cite{Mandula:2007jt,Mandula:2009yd}.

A linear transformation $\psi\to e^{i\varphi X}\psi$, $\bar\psi\to\bar\psi\,e^{i\varphi\bar X}$ leaves the action invariant if and only if $\bar XD+DX=0$. The three symmetries U(1)$_V$, U(1)$_A$, $\widetilde{\mathrm{U(1)}}_A$ correspond to
\begin{equation}
(X,\bar X)=(1,-1),\qquad (\hg,\g_5),\qquad (\g_5,(1-D)\g_5),
\end{equation}
respectively, where the last two follow from $D\hg=-\g_5D$ and from the Ginsparg--Wilson relation in the form $(1-D)\g_5D=-D\g_5$.

The operator $V=1-D$ is unitary and satisfies $\g_5V\g_5=V^\dagger$. Following \cite{Mandula:2007jt,Mandula:2009yd,Kourkoulou:2026lzx}, we define
\begin{equation}
Q_n=\g_5V^n,\qquad G_n=-i\,(V^n-V^{-n}),\qquad n\in\mathbb Z .
\label{eq:QG}
\end{equation}
The $Q_n$ are hermitian and square to one, and the $G_n$ are hermitian. Using $Q_nQ_m=V^{m-n}$ one finds
\begin{equation}
[Q_n,Q_m]=i\,G_{m-n},\qquad [Q_n,G_m]=2i\,(Q_{n-m}-Q_{n+m}),\qquad [G_n,G_m]=0 ,
\label{eq:onsager}
\end{equation}
which is the Onsager algebra. Our two axial generators are $Q_0=\g_5$ and $Q_1=\hg$, and $Q_{-1}=(1-D)\g_5$. Since $Q_0$ and $Q_1$ are involutions, they satisfy the Dolan--Grady relations
\begin{equation}
\big[Q_1,[Q_1,[Q_1,Q_0]]\big]=4\,[Q_1,Q_0],\qquad \big[Q_0,[Q_0,[Q_0,Q_1]]\big]=4\,[Q_0,Q_1],
\end{equation}
and generate the whole algebra \eqref{eq:onsager}.

Each element of \eqref{eq:onsager} generates a symmetry of the overlap Dirac fermion. Using $D=1-V$ and $D\g_5=\g_5(1-V^{-1})$, one checks that $DQ_n=-Q_{n-1}D$ and $DG_n=G_nD$. Hence
\begin{equation}
\begin{aligned}
&R_{Q_n}(\varphi):\quad \psi\to e^{i\varphi Q_n}\psi,\quad \bar\psi\to\bar\psi\,e^{i\varphi Q_{n-1}}\,,\\
&R_{G_n}(\varphi):\quad 
\psi\to e^{i\varphi G_n}\psi,\quad \bar\psi\to\bar\psi\,e^{-i\varphi G_n}\,,
\label{eq:onsagersym}
\end{aligned}
\end{equation}
leave the action invariant. 
The transformations associated with $Q_n$ with $n=1$ and $n=0$ correspond to $\mathrm{U(1)}_A$ and its reflection $\widetilde{\mathrm{U(1)}}_A$, respectively, i.e., $R_{Q_1}=R_A$ and $R_{Q_0}=\tilde R_A$. The Grassmann measure is also invariant, since the $Q_n$ are traceless and the $G_n$ act with opposite signs on $\psi$ and $\bar\psi$. 
Every $Q_n$ squares to one, so $R_{Q_n}(\varphi)$ is $2\pi$-periodic and  generates a compact $\mathrm{U(1)}$ symmetry. 
Note that  $e^{i\pi Q_n}=-1$ is the fermion parity for every $n$. 
By contrast, the transformation $R_{G_n}(\varphi)$ is not periodic in $\varphi$ and generates a  noncompact symmetry. Denoting by $\mathbf{Ons}$ the group generated by \eqref{eq:onsagersym}, the symmetry of the overlap Dirac fermion generated by the three $\mathrm{U(1)}$s is
\begin{equation}
\frac{\mathrm{U(1)}_V\times\mathbf{Ons}}{\mathbb Z_2},
\end{equation}
where $\mathrm{U(1)}_V$ commutes with $\mathbf{Ons}$, and the $\mathbb Z_2$ identifies the fermion parity in $\mathrm{U(1)}_V$ with that in $\mathbf{Ons}$.

The antilinear Dirac reflection $\TD$ acts on $\mathbf{Ons}$ as an automorphism. Following the calculation that led to \eqref{eq:mirrorA}, one finds that $\TD$ acts on the symmetry transformations \eqref{eq:onsagersym} as
\begin{equation}
\TD R_{Q_n}(\varphi) \TD= R_{Q_{1-n}}(\varphi)\,,\quad
\TD R_{G_n}(\varphi) \TD =R_{G_n}(-\varphi)\,,
\end{equation}
where we have used $V^{\TD}=1-D^\dagger=V^{-1}$. 
No $R_{Q_n}(\varphi)$ is mapped to itself, so none of the compact axial $\mathrm{U(1)}$s in $\mathbf{Ons}$ is invariant under the reflection. 
The reflection-invariant combination $\frac12(Q_0+Q_1)=\g_5(1-\frac12D)$ generates the noncompact $\mathbb{R}$ symmetry \eqref{eq:qA}. 
These observations are tied to the fact that in the Hamiltonian formulation of the overlap Dirac fermion model, the analogous Onsager algebra does not commute with the Hamiltonian \cite{Kourkoulou:2026lzx}, while the axial symmetry is noncompact \cite{Creutz:2001wp,Singh:2025sye}.

Finally, in the continuum limit $D\to0$, every $Q_n$ reduces to the continuum axial generator $\g_5$ and every $G_n$ vanishes. The symmetry $[\mathrm{U(1)}_V\times\mathbf{Ons}]/\mathbb Z_2$ therefore reduces to $[\mathrm{U(1)}_V\times \mathrm{U(1)}_A]/\mathbb Z_2$, which is a subgroup of the entire continuous internal global symmetry of a free massless Dirac fermion.

 \section{Torus partition function of the overlap Weyl fermion}\label{sec:Z}

Despite the conceptual aspects of the overlap Weyl fermion discussed in the main text, it is instructive to set them aside and perform an explicit calculation. 
The partition function of the overlap Weyl fermion on a two-dimensional torus with twisted boundary conditions was computed in \cite{Narayanan:1994gb,Fosco:1995pu,Fosco:1995rc}, where it was shown to agree with the continuum result. 
In this appendix we review this calculation, and match the modulus of the partition function with the continuum result in conformal field theory in the thermodynamic limit.

In $d=2$ the chiral projection can be solved explicitly with the local sections of the Weyl bundle, and the partition function can be computed exactly. 
We consider a general torus, obtained from $\mathbb Z^2$ by the identifications $x\sim x+w_1$ and $x\sim x+w_2$ with
\begin{equation}
w_1=(L_1,0),\qquad w_2=(s,L_2),\qquad L_1,L_2,s\in\mathbb Z ,
\end{equation}
so that it has $V=L_1L_2$ sites and modular parameter
\begin{equation}
\tau=\tau_1+i\tau_2=\frac{s+iL_2}{L_1}.
\end{equation}
The rectangular torus of the main text is $s=0$.
We impose the twisted boundary conditions ($\theta_\mu\in [0,1)$)
\begin{equation}
    \psi(x+w_\mu)=e^{2\pi i\theta_\mu}\psi(x)\,,\qquad \mu=1,2,
\end{equation}
so that $\theta_\mu=0$ ($\theta_\mu=\frac12$) corresponds to the Ramond (Neveu--Schwarz) sector in the $\mu$ direction. 
The allowed momenta obey $p\cdot w_\mu=2\pi(n_\mu+\theta_\mu)$, i.e.,
\begin{equation}
p_1=\frac{2\pi}{L_1}(n_1+\theta_1),\qquad p_2=\frac{2\pi(n_2+\theta_2)-s\,p_1}{L_2},\qquad n_\mu=0,1,\dots,L_\mu-1 .
\end{equation}

We expand the fermion field in momentum space as
\begin{equation}
\psi(x)=\frac1{\sqrt V}\sum_pe^{ip\cdot x}\,u_{\rm N}(p)\,c_p,\qquad \bar\psi(x)=\frac1{\sqrt V}\sum_pe^{-ip\cdot x}\,\bar c_p\,(1,0),
\end{equation}
with $u_{\rm S}$ in place of $u_{\rm N}$ at the would-be doublers $(\pi,0),(0,\pi),(\pi,\pi)$, whenever they are allowed momenta. 
Recall that $u_\text{N,S}(p)$, defined in \eqref{eq:charts}, are solutions to the chiral projection $\hat P_-(p)u_\text{N,S}(p) = u_\text{N,S}(p)$. 
Since $D(p)u_{\rm N}(p)$ lies in $\im P_+$, the action is diagonal,
\begin{equation}
S=\sum_p\bar c_p\,M(p)\,c_p,\qquad M(p)=i(\beta_1-i\beta_2)\sqrt{\frac{2}{1+\alpha}},\qquad |M(p)|^2=2(1-\alpha),
\label{eq:M}
\end{equation}
where in $d=2$,
\begin{equation}
\alpha(p)=\frac{\cos p_1+\cos p_2-1}{\omega(p)},\qquad
\beta_\mu(p) = {\sin p_\mu \over \omega(p)}, \qquad
\omega(p)^2=1+2(1-\cos p_1)(1-\cos p_2).
\end{equation}
Near $p=0$, $M\simeq i(p_1-ip_2)$ is the continuum Weyl kernel, and at the doublers $|M|=2$. 
The squared modulus of the partition function is
\begin{equation}
|Z|^2=\prod_p2\big(1-\alpha(p)\big)\quad
\Leftrightarrow
\quad 
2 \ln |Z| = \sum_p \ln [2(1-\alpha(p))],
\end{equation}
whereas the phase is ambiguous because of the freedom to rephase the local sections.
Indeed, changing $u_{\rm N}(p)\to e^{i\varphi_p}u_{\rm N}(p)$ multiplies $Z$ by $\prod_pe^{-i\varphi_p}$, which is L\"uscher's phase ambiguity of Section~\ref{sec:weyl}.

The function $\ln[2(1-\alpha(p))]$ is $2\pi$-periodic in $p_1$ and $p_2$, so we can trivially rewrite it as
\begin{equation}
\ln\big[2(1-\alpha(p))\big]=\sum_{x\in\mathbb Z^2}e^{ip\cdot x}\int\frac{d^2k}{(2\pi)^2}\ln\big[2(1-\alpha(k))\big]\,e^{-ik\cdot x},
\end{equation}
with $x=(x_1,x_2)\in\mathbb Z^2$ a lattice site. 
Summing over the allowed momenta $p$ now gives
\begin{equation}
\sum_pe^{ip\cdot x}=V\sum_{m_1,m_2\in\mathbb Z}e^{2\pi i(m_1\theta_1+m_2\theta_2)}\,\delta_{x,\,m_1w_1+m_2w_2}.
\end{equation}
Hence, using the above two equations and renaming $k$ as $p$, we have
\begin{equation}
\ln|Z|^2=V\sum_{m_1,m_2\in\mathbb Z}e^{2\pi i(m_1\theta_1+m_2\theta_2)}\int\frac{d^2p}{(2\pi)^2}\,\ln\big[2(1-\alpha(p))\big]\,e^{-ip\cdot x}\Big|_{x=m_1w_1+m_2w_2} .
\label{eq:Zwind}
\end{equation}
So far this is an exact formula without approximation. 
The term $m_1=m_2=0$ is $-fV$, where $-f$ is the Brillouin-zone average of $\ln[2(1-\alpha(p))]$,
\begin{equation}
-f= \int\frac{d^2p}{(2\pi)^2}\ln\big[2(1-\alpha(p))\big]=0.951668\ldots .
\label{eq:f}
\end{equation}
The other terms are the error made by replacing the integral \eqref{eq:f} by a sum over discrete momenta. 

The large-volume behavior of these terms is controlled by the large-distance behavior of the Fourier coefficients. The Fourier coefficients of a smooth periodic function decay exponentially in $|x|$. Here the function $\ln[2(1-\alpha(p))]$ is smooth except at $p=0$, where it equals $\ln|p|^2$ plus terms of order $p^2$.
Since $\nabla_p^2\ln|p|^2=4\pi\delta^2(p)$, integrating by parts twice gives $-|x|^2\int\frac{d^2p}{(2\pi)^2}\ln|p|^2e^{-ip\cdot x}=\frac1\pi$. For large $|x|$, we therefore have 
\begin{equation}
\int\frac{d^2p}{(2\pi)^2}\,\ln\big[2(1-\alpha(p))\big]\,e^{-ip\cdot x}=-\frac{1}{\pi|x|^2}+O(|x|^{-4}).
\end{equation} 
The $O(|x|^{-4})$ terms contribute $V\sum_{(m_1,m_2)\neq(0,0)}O(|m_1w_1+m_2w_2|^{-4})=O(L^{-2})$ to \eqref{eq:Zwind}. Using $|m_1w_1+m_2w_2|^2=L_1^2|m_1+m_2\tau|^2$ and $V/L_1^2=\tau_2$, we find
\begin{equation}
\ln|Z|^2+fV=-\sum_{(m_1,m_2)\neq(0,0)}\frac{\tau_2}{\pi|m_1+m_2\tau|^2}\,e^{2\pi i(m_1\theta_1+m_2\theta_2)}+O(L^{-2}).
\label{eq:Eisenstein}
\end{equation}
The right-hand side is related to the Jacobi theta function  by Kronecker's second limit formula (see Chapter~1, \S3 of \cite{Siegel:1961lect} and also \cite{DHoker:2015foa} for a physics application):\footnote{The identity holds for $\theta_1,\theta_2$ not both zero. When both vanish, $Z=0$ because of the fermion zero mode at $p=0$.}
\begin{equation}
\sum_{(m_1,m_2)\neq(0,0)}\frac{\tau_2}{\pi|m_1+m_2\tau|^2}\,e^{2\pi i(m_1\theta_1+m_2\theta_2)}=-2\ln\left|\frac{\vartheta[\theta_1-\frac12,\frac12-\theta_2](\tau)}{\eta(\tau)}\right| ,
\label{eq:Kronecker}
\end{equation}
where
\begin{equation}
\vartheta[a,b](\tau)=\sum_{n\in\mathbb Z}e^{i\pi\tau(n+a)^2+2\pi i(n+a)b},\qquad \eta(\tau)=q^{1/24}\prod_{j\ge1}(1-q^j),\qquad q=e^{2\pi i\tau}.
\end{equation}
Combining \eqref{eq:Eisenstein} and \eqref{eq:Kronecker}, we conclude
\begin{equation}
\ln|Z|^2=-fV+2\ln\left|\frac{\vartheta[\theta_1-\frac12,\frac12-\theta_2](\tau)}{\eta(\tau)}\right|+O(L^{-2}).
\label{eq:ZL}
\end{equation}
Up to the nonuniversal bulk term $-fV$, \eqref{eq:ZL} is the squared modulus of the partition function of a continuum Weyl fermion on a torus of modulus $\tau$ with twisted boundary conditions (see eqs.~(2.1)--(2.3) of \cite{Alvarez-Gaume:1986rcs}). In particular, for $(\theta_1,\theta_2)=(\frac12,\frac12),(\frac12,0),(0,\frac12)$ we reproduce
\begin{equation}
\big|\mathrm{Tr}_{\rm NS}\,q^{L_0-\frac1{24}}\big|^2=\Big|\frac{\vartheta_3}{\eta}\Big|^2,\qquad
\big|\mathrm{Tr}_{\rm NS}\,(-1)^Fq^{L_0-\frac1{24}}\big|^2=\Big|\frac{\vartheta_4}{\eta}\Big|^2,\qquad
\big|\mathrm{Tr}_{\rm R}\,q^{L_0-\frac1{24}}\big|^2=\Big|\frac{\vartheta_2}{\eta}\Big|^2,
\end{equation}
respectively, where the traces are over the Hilbert space of the continuum Weyl fermion quantized on the cycle $w_1$.

\bibliographystyle{JHEP}
\bibliography{ref}

\end{document}